\documentstyle[aps, preprint]{revtex}
\begin{document}
\draft
\title{\bf{Weakly localised bosons}}
\author{D. Das and S. Doniach\\Department of Applied Physics\\
	Stanford University, Stanford, CA 94305. U.S.A.}
\date{Revised Draft 21 Jan 98}

\maketitle
\begin{abstract}
In this paper, we argue that, in 2-d, the weak localisation of bosons, which
occurs on the insulating side of the superconductor-insulator transition, is
characterised by $\rho\sim ln(1/T)$, as compared to $\sigma\sim lnT$ for
fermions. Such an unconventional behaviour is tied to the diffusion pole
of the delocalised (uncondensed) vortices. \\

\end{abstract}
\newpage

Superconductivity in low-dimensional systems has been of considerable
interest over the past few decades. This has received a strong boost
with the discovery of high-$T_{c}$ superconductivity. High-$T_{c}$
superconductors are layered materials and exhibit a
superconductor-insulator transition at low doping. In an attempt to
understand the two-dimensional aspect of this new phenomenon, an
appreciable amount of effort has been focussed on the low-temperature
behaviour of thin films made out of the low-$T_{c}$
superconductors[1].  In these experiments, one observes that the film
undergoes a transition from a superconductor to an insulator at
$T\rightarrow 0$, as one tunes the thickness, disorder or the magnetic
field[1]. There are two distinctively different schools of thought,
for explaining this set of observations. One school [2] believes that the
insulating behaviour occurs due to the localisation of charged bosons,
viz. Cooper pairs. Whereas, the other school[3] believes that during
the said transition, Cooper pairs break up into fermions which
naturally localise in 2-d and cause insulating behaviour. Each school
is supported by a corresponding theory. Cooper pair localisation is
supported by Fisher's theory[4], whereas the fermion localisation
picture is interpreted in terms of Fukuyama's theory [5]. In the
latter scenario, Cooper pairing is hampered as a result of suppression
of density of states due to enhanced coulomb repulsion. Although
Fisher's theory seemed to work in many cases[6], there has been
unflinching evidence[3,7] that Fukuyama's theory is right for certain
systems. Thus each mechanism probably has an appropriate class of
materials for which it is relevant.

So far, there has been no straightforward way to find which theory
applies in a given system. In this paper, we suggest a simple way to
resolve this debate. In particular, we argue that, in 2-d, weak
localisation of bosons implies $\rho\sim ln(1/T)$, whereas for fermion
localization one expects $\sigma\sim lnT$ [8].
The closest experimental observation of this non-fermionic effect seems 
to be the
experiment by Gerber on granular Pb films[9].
 A similar behaviour has been reported in Ref.[6]. It is interesting to note 
that such a $\rho\sim ln(1/T)$ dependence has been observed in the high-$T_{c}$
materials[10] as well, where bosonic models have been suggested to be relevant[11].
In what follows, most of our discussion will be restricted to two
dimensional systems at finite temperatures, the focus being exclusively on
low-$T_{c}$ superconductors. 

{\em Why not $\sigma\sim lnT$ for bosons?} $\sigma\sim lnT$ behaviour
is profuse in 2-d electronic systems. Many of these logarithmic
tendencies are associated with the critical properties of electrons,
viz. the lower critical dimension of the electronic system is two. By
contrast, in a bosonic system, the lower critical dimension is one[4].
There is a further difference between the two kinds of charge
carriers. Let us consider, for example, a system of non-interacting
electrons. Here $\sigma\sim lnT$ follows from one-parameter
scaling[8]. It has been argued that [12] the scaled conductance
$g=G/(e^{2}/\hbar)$ for such systems depends on a single parameter
$\Delta E/\delta W$, where $\Delta E$ = bandwidth and $\delta W =
(N_{0}L^{d})^{-1}$ is the scatter in the energy levels in a system of
size L in d-dimensions. $N_{0}$ is the density of states at the Fermi
level. Let us repeat the argument[12] here for the sake of
completeness. In this picture, one considers constructing a sample of
size $(2L)^{d}$, in $d$ dimensions, out of $2^{d}$ samples of size
$L^{d}$. A given eigenstate of the larger sample is a linear
combination of the eigenstates of the smaller samples. What really
matters here is the amount of admixture among the (old) eigenstates,
which in turn is controlled by the overlap integral and energy
denominator.  The measure of the energy denominator is $\delta W$. To
obtain an estimate for the overlap, one considers repeating a given
$L^{d}$ sample in one direction, subject to appropriate boundary
conditions. Each eigenstate broadens to form a band, and the bandwidth
$\Delta E$ is an estimate of the overlap integral. If the wave
function is localised, $\Delta E$, and hence, ${\Delta E}/{\delta W}$,
is exponentially small. If the wavefunction is delocalised, ${\Delta
E}/{\delta W}$ is large and $\Delta E$ is sensitive to boundary
conditions. Thus, the scaled conductance $g$ is a simple function of a
single parameter ${\Delta E}/{\delta W}$.  But this is not enough for
a bosonic system. The phase of the wavefunction is an important
parameter here. A pretty argument due to Anderson[13] suggests that,
for the Bose case, there are strong phase relationships among the
(smaller) samples, discussed above, which would invalidate
one-parameter scaling and also the logarithmic behaviour of the
conductivity along with it. More precisely, in the localised phase,
the so called Bose-glass(BG) phase[14], there are big patches of
locally superfluid regions which strongly
modify the scattering of the charge bosons in the following way.
When a charge boson scatters off an impurity, it correlates with(or, talks to)
all the charge bosons within a distance of the order of the localisation 
length $\xi_{loc}$. As a 
result, the boson scatters as a blob, i.e. the scattering is inherently
multiparticle/collective in nature, as compared to a single particle 
scattering typical of a non-interacting particle. (Plus, this is also the 
physical reason why we find a tenfold enhancement of resistance over the
fermionic counterpart.)
 That is why it
is much more convenient to think in terms of the collective excitations, viz.
vortices, which are dual to the
charge (boson) degrees of freedom[15]. Translated into the vortex picture, 
big superconducting patches convert into regions where vortices are localised.
In between these large regions, there are {\em narrow} channels where the 
vortices are delocalised, and only these make a significant contribution
 to the vortex conductivity[16]. These channels are superfluid below a certain 
temperature $T_{\lambda\rm V}$(please see later), above which they lose phase
coherence[17].
Further, we note that[18] the charge
resistivity
$\rho_{c}$ is related to the vortex conductivity $\sigma_{v}$ by
\begin{equation}
   \rho_{c} = (h/4e^{2})\sigma_{v}
\end{equation}

In the rest of the paper, we are going to exploit this duality relation
between charges and vortices extensively and evaluate the resistance as seen
by the charges in terms of the vortex mobility.

In order to calculate the vortex conductivity, or equivalently the
charge resistivity, we consider a model of interacting (charged) bosons in
a random potential[14]. Before we write down this model explicitly,
let us discuss the physical picture[Fig.1] generated by this model. 
Consider a
system of bosons(e.g. Cooper pairs in a granular superconductor)
interacting with a short-range repulsive potential $V$ in the presence
of a finite disorder $\Delta$. Let $J$ be the zero point energy of the
bosons. When the ratio $(J/V)$ is large, the bosons condense into a
superfluid(SF) state. As the ratio $(J/V)$ is decreased (or, alternatively,
$(\Delta/V)$ is increased), the bosons tend to localise
and superfluidity decreases. At a critical value of the kinetic energy
$(J/V)_{c}$, (global) superfluidity is completely destroyed and the bosons
make a transition to the Bose-glass phase, where the bosons are essentially
localised by disorder. On the insulating side, very close to the
phase boundary, the bosons are weakly localised and there are finite
regions (of the order of the localisation length $\xi_{loc}$) where the
bosons form a local condensate. As 
one moves further from the phase boundary by reducing $(J/V)$, one
gradually enters the strongly localised regime. Here the overlap between
the nearest neighbours is really small and Mott variable range hopping
is expected to occur[18]. In this paper, we address
the issue of conductivity of Cooper pairs in the Bose glass phase
very close to the BG-SF phase boundary. For the sake of completeness, we
note that when the average number of
bosons per site is an integer, there is an additional Bose insulating(BI)
phase. In this phase, there are exactly n bosons per site (where n is an
integer) and the phase is gapped.

One can consider this entire phase diagram in the (dual) vortex
picture as well. In the superfluid region, the vortices form closed
loops in zero field.  At the superfluid-Bose glass transition, the loops blow up as
a result of the quantum zero point motion at $T=0$ and thermal motion
at finite $T$, and the vortex-antivortex pairs break up[19]. 
 Since these vortices and antivortices are bosons, they
 form a bose condensate at $T=0$[15]. However, for $T >
T_{\lambda\rm V}$, the lambda transition of the superfluid vortex
condensate[20],
 this is destroyed, because
the disordering of the charges(viz. Cooper pairs for this case)[18],
which represent the topological excitation over the vortex degrees of
freedom, will destroy any possible phase coherence among the
vortices. We further note that the existence of a vortex condensate
would simply mean, from (1) that, $\rho_{c}=\infty$ at a finite
temperature. This signature of a ``quantum insulator'' is rather
counterintuitive, so far as the resistive behaviour of classical
insulators is concerned. However, it can be made clearer in the dual 
``cooper lattice'' phase where the system is essentially a charge density 
wave in which the pairs are strongly correlated. The superinsulator 
behaviour then corresponds to the response of a sliding charge density wave
in the absence of pinning. This behaviour would probably be suppressed
experimentally by finite size and nonlinear effects. In the rest
of the paper, we shall work in the regime $T >
T_{\lambda\rm V}$ which we assume may be quite small in the materials
such as those studied by Gerber.

Although the above phase
diagram is simplest to  model on a lattice, in the rest of this paper
we shall consider the bosons in a continuum[21,15,22].
 The main purpose of this exercise is to
demonstrate that under the duality transformation, vortices see
potential disorder(as seen by the charges) as a random magnetic field(RMF).
This feature is particularly explicit in a continuum description. Hence
we consider the following model Hamiltonian$(\hbar=1)$
\begin{equation}
   H = \int d^{2}x[\frac{1}{2m}\mid\frac{1}{i}\vec{\nabla}\psi\mid^{2} +
\frac{1}{2}\int d^{2}y\delta\rho(x)V(x-y)\delta\rho(y) -
\mu(\vec{x})\rho(\vec{x},\tau)],   
\end{equation}
where $\delta\rho=\rho-\bar{\rho}$, $\bar{\rho}$ being the (neutralising)
background charge density with partition function $Z=\int{\mathcal D}\psi^{\ast}{\mathcal
D}\psi exp(-S)$, and $S=\int d\tau d^{2}x{\mathcal L} = \int d\tau
[\int d^{2}x \psi^{\ast}\frac{\partial\psi}{\partial\tau}+H]$. Here
$V(x)$ is a short-range potential $V(x)=V\delta(x)$. 

Equivalence with the above phase diagram is established by setting
$\bar{\rho}/m = J$. Now, invoke a duality transformation[23]
$\psi=\sqrt{\rho}e^{i\theta}\phi_{v}$, where $\theta=$
non-topological part of the phase and $\phi_{v}=$topological part of the
wave-function, with $\phi_{v}^{\ast}\phi_{v}=1$. Also,
$\mu=\mu_{0}+\delta\mu_{i}(\vec{x})$, where $\delta\mu_{i}(\vec{x})$ represents the
impurity potential which is Gaussian distributed with variance $\Delta$.
(We have not written down the chemical potential $\mu_{0}$ explicitly in
what follows for convenience, and have taken $\mu(\vec{x})$ as
$\delta\mu_{i}(\vec{x})$.)
Further, define a gauge field $M_{i}(\vec{x})$ ($i=1,2$) by the
relation
$\epsilon_{ij}\partial_{i}M_{j}=-\mu(\vec{x})/V\equiv B(\vec{x})$. Here
$\vec{M}(\vec{x})$ is independent of $\tau$. As a result, one is led to
the following Lagrangian --
\begin{equation}
   {\mathcal L}=\frac{m}{2\bar{\rho}}\mid\vec{J}\mid^{2} + \frac{V}{2}(\delta\rho)^{2} +
2\pi i \tilde{j_{\mu}}a_{\mu} + 2\pi i \tilde{j_{k}}M_{k}, 
\end{equation}
$(\mu = 0,1,2, k=1,2) $, with 
\[ <B(\vec{x})B(\vec{x'})> =
(\Delta/V)^{2}\delta(\vec{x}-\vec{x'}). \]
Here
$J_{\mu}=(\delta\rho,\vec{J})=\epsilon_{\mu\nu\rho}\partial_{\nu}a_{\rho}$
represents the charge (boson) density and current, whereas
$\tilde{j_{\mu}}=j_{\mu}^{v}=(1/2\pi
i)\epsilon_{\mu\nu\rho}\partial_{\nu}(\phi_{v}^{\ast}\partial_{\rho}\phi_{v})$
represents the vortex density and current[24]. Integrating out the charge
degrees of freedom, and transforming over to the real time, we are led to the following model Hamiltonian 
\begin{equation}
   H_{v} = \sum_{i\alpha}\frac{1}{2m_{v}}(p_{i\alpha}-2\pi
q_{i}M_{i\alpha})^{2} +
\frac{1}{2} 4\pi^{2}\frac{\bar{\rho}}{m} \sum_{i\neq j} q_{i}q_{j} ln(\mid
x_{i}-x_{j}\mid/\xi_{0}) + consts   
\end{equation}
Here $q_{i}=\pm 1,(\alpha=1,2)$ represents the charge on a vortex and
an antivortex respectively, $x_{i}$ and $p_{i}$ represent the
positions and momenta of vortices and summation is over all vortices
and antivortices. Here $m_{v}$ refers to the vortex mass. Details of
the notation and missing steps can be found in [23]. Thus, we find
that vortices move in a random effective magnetic field under a duality
transformation. A few comments are in order. In going from eqn (3) to
(4), we have included a vortex mass term. This kinetic energy term is
essential for the quantum melting of the 
 vortices and is guaranteed by the underlying electronic
degrees of freedom[25].
Also, we have neglected a current-current
interaction term among vortices resulting from the short-range
interaction among charges.  This is valid when $n_{v}/Vm_{v}\ll
1/4\pi^{2} (n_{v}=$vortex density) and at low temperatures. The
physical content of this condition is as follows. The mass term for
vortices generates a long-range interaction among the charges[26]. The
above inequality simply states that this long-range component is much
weaker than the short-range component.  Thus, the BG phase discussed
above is perturbed to as little extent as possible.

As we noted earlier, in the BG phase, the vortex-antivortex pairs break
up. Since they are delocalised, they screen each other, and as a result,
interact via short-range interactions. Thus, we are led to evaluate the
conductivity of a vortex liquid as it diffuses in the presence of a
random magnetic field at finite temperature. We shall do this using
perturbation theory. As is well known in the
context of the fermionic problem[27,28], the diffusion pole,
viz.$\Gamma(q,\omega_{n}) =
({1}/{m\tau^{2}})/(\mid\omega_{n}\mid + Dq^{2})$, for 
$\epsilon_{m}(\epsilon_{m}+\omega_{n})<0$, (where $\omega_{n}=2\pi nT$, $n=$integer, are
Matsubara frequencies and $D$ is the diffusion constant) generates
singular corrections to the conductivity(at finite temperature). In the
absence of a condensate, this is true of the Bose case as well, with
$m=m_{v}, D=\frac{1}{2}v_{b}^{2}\tau_{tr}$, where
$\frac{1}{2}m_{v}v_{b}^{2}=\mu_{v}\equiv E_{b}=$chemical potential of vortices,
$\tau_{tr}=$transport time in a
RMF$=m_{v}\xi_{0}^{2}/{\pi^{2}(\Delta/V)^{2}}$, and $\xi_{0}=$pair
size, defines the microscopic scale of the Bose problem.
$\tau=$ the elastic scattering time in the RMF, and we will take
$\tau=\tau_{tr}$[28]. Now, for superconductors,
$\mu_{v}=d(\Phi_{0}/4\pi\lambda)^{2}ln\kappa$, where $\kappa$ is
the Ginzburg-Landau parameter and $d$ is the film thickness.  One can rewrite this as
$\mu_{v}=\alpha_{0}\bar{\rho}/m\equiv\alpha_{0}J$, with
$\alpha_{0}=\frac{\pi}{4}ln\kappa$. Using the
usual value for the Josephson coupling energy $J=(R_{Q}/2R_{n})\Delta_{0}$,
where $\Delta_{0}$ is the superconducting gap,
$R_{Q}=h/4e^2\simeq6.45K\Omega$, and $R_{n}$ is the normal
resistance of the sample, we obtain the condition for
validity of perturbation theory as 
\begin{equation}
   E_{b}\tau_{tr} =
\frac{2}{\pi}\alpha_{0}\frac{k_{F}d}{(\Delta/V)^{2}}\frac{E_{F}}{\Delta_{0}}\frac{R_{Q}}{R_{n}} \gg 1 
\end{equation}
Because magnetic disorder is time-reversal symmetry breaking, the
cooperon mode is suppressed[28]. The diagrams which contribute are the
Altshuler-Aronov type diagrams, shown in fig.2. 
We merely quote the result here and refer the reader to
Refs.  [27,28] for more details. We obtain
\begin{equation}
    \delta\sigma_{v} = -A_{0}ln(T\tau_{tr})  
\end{equation}
an extra minus sign coming from the replacement of anticommutators by
commutators for bosons. Physically, this makes sense because as we go
down in temperature the vortices being bosons tend to bose-condense.
Here $A_{0}=(2/\pi)(1+F/2)$, a number of order unity, with
$F=\frac{1}{2}/\sqrt{1+(\alpha_{0}/2\pi^{2})}.$
Eqn.(6) is an
enhancement over a background part
$R_{v0}\mid_{RMF}=(h/4e^{2})(n\tau_{tr}/m_{v})\sim(h/4e^{2})
(n_{v}\xi_{0}^{2}/(\Delta/V)^{2})$.
The constant part of the conductivity will receive contributions from the
Bardeen-Stephen processes, $R_{BS}=2\pi R_{n}(n_{v}\xi_{0}^{2})$[29]. Putting all
these contributions together, we obtain,
\begin{equation}
 \rho_{c} = R_{b} + R_{0}ln(T_{0B}/T), 
\end{equation}
where $R_{b}=[2\pi
R_{n}+(R_{Q}/(\Delta/V)^{2})](n_{v}\xi_{0}^{2})$, and $R_{0}=A_{0}R_{Q}$. Also, in
the BG phase, we expect $(\Delta/V)^{2}$ to be a number of order unity. Here
$T_{0B}=1/\tau_{tr}$. Eqn.(7) is the main result of the paper.

{\em Comparison with experiment.} The logarithmic behaviour for bosons,
as predicted by eqns.(6) and (7), is much steeper than the analogous fermionic
behaviour (with interactions included), viz.
$\sigma_{F}=\sigma_{0}-(2-2ln2)(e^{2}/\pi h) ln(T_{0F}/T)$ [12],
with $\sigma_{0}=(e^{2}/h)(k_{F}l)$. For fairly large values of
$k_{F}l$, this implies
\begin{equation}
   \rho_{F}\simeq\rho_{0} + R_{F}ln(T_{0F}/T), 
\end{equation}
where $1/T_{0F}=\tau_{F}=$electronic transport time, and
$R_{F}=(2-2ln2)(\rho_{0}/\pi k_{F}l)$.  
Since we do not have all the
necessary 
material parameters available for Gerber's experiment[9], we make some estimates based on
Mo-Ge samples discussed in Ref.[6]. In this case, 
$T_{\lambda\rm V}\sim10mK$[20], much lower than the temperature range 
of measurements. This justifies our neglect of the ``quantum insulator'' phase
in this discussion.
Taking $k_{F}l\sim5$[30],
we have $R_{F}\simeq0.1-1K\Omega$. The scale of $R_{0}$ is set by $R_{Q}=
(h/4e^{2})$
and we have $R_{0}\sim5k\Omega$.
Thus, we expect a change in slope of a $R$ vs $lnT$ curve by a factor
of about 10 when bosonic conduction sets in.  There is one more
difference between Bose and Fermi resistivities. In the case of the
former, $T_{0B}\simeq 0.1-10K$, whereas $T_{0F}\simeq
10^{2}-10^{3}K$. Thus, the fermionic mechanism is a high temperature
phenomenon, whereas the bosonic behaviour is a predominantly low
temperature phenomenon. For low-$T_{c}$ materials, $T_{0B}
\sim T_{p}$, the pair formation temperature, and the 
$lnT$ dependence may be limited by $T_{p}$ and set in
as soon as the pairs are formed. These features are clearly observable 
in Gerber's experiment[9]. Also, in this experiment, at a very high
magnetic field which kills quasi-reentrance (characteristic of pair
formation), this low temperature logarithmic behaviour disappears as
well. We consider this to be a telltale evidence 
of bosonic transport. However, it must be mentioned that in this
experiment[9],
the scale of the resistivity is unusually high, lying between megaohms
and
gigaohms. For a typical resistivity of fig.2 in Ref.[9], we get
$R_{Q}/R_{n}\sim10^{-4}$. This means, from eqn(5), that
$E_{b}\tau_{tr}\sim O(1)$. So, higher order corrections in
$(1/E_{b}\tau_{tr})$
are particularly important in this case and probably a certain class of
diagrams needs to be resummed.
We also take this opportunity to comment that Wolf's observations[31],
which originate on the high temperature side, are most likely not due
to bosonic transport, but have some fermionic mechanism attached to
it. 

Our focus in this paper has been only on short range interactions among 
the charge 
bosons, although in the superconducting films like Mo-Ge[6], which are
of current interest, long-range interactions dominate. It is quite possible
 that 
inclusion of such effects will renormalise the coefficients only, rather
than affecting the temperature dependence strongly. This needs further 
investigation.

All these issues prevent us from making good contact with the existing
experiments and we only hope to have conveyed to the reader the essence
of bosonic transport, and how it is distinguished from its fermionic analog.
 We believe, however, that this behaviour should be seen in other
systems where bosonic transport dominates.

{\em What this is not?} We would like to comment here, before closing, 
that this model is {\em not} the same as the traditional Bose Hubbard 
model with simple on-site repulsion. The latter does not generate a 
vortex mass term[32] which is 
particularly important for vortex mobility calculations discussed above. 
Thus, any simulation which tries to track this resistive behaviour must
include an appropriate vortex mass term along with the usual Bose Hubbard model
terms.

To summarise, we have argued that in the regime where Cooper pairs are
weakly localised, the vortices are in a liquid phase. Thermal
diffusion of this quantum liquid in the presence of (pseudo-magnetic)
disorder leads to logarithmic temperature dependence of the (charge)
resistivity. This kind of temperature dependence should be observable
in a bosonic system, so long as the temperature is higher than that at
which the vortices themselves would form a bosonic condensate.

\newpage
\begin{figure}
\caption{A schematic phase diagram for the Bose localisation problem. The
hatched region shows the region of interest.}
\caption{Altshuler-Aronov diagrams}
\end{figure}

\begin{thebibliography}{99}
\bibitem{cm4} The view of the weakly localised phase 
  taken here is, in a way, that of large regions of charge
condensate getting pinned by disorder.(Please see Ref.22 for more details.)
\bibitem{cm} Lattice model has an extra Bose insulating phase.
\bibitem{cm2} In this we have neglected a constant magnetic field(seen
by vortices) due to a background charge condensate. This is appropriate
in a BG phase.
\end{thebibliography}
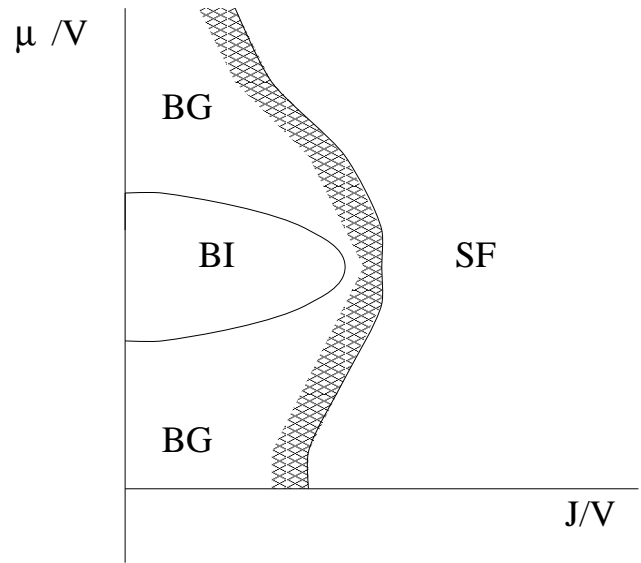
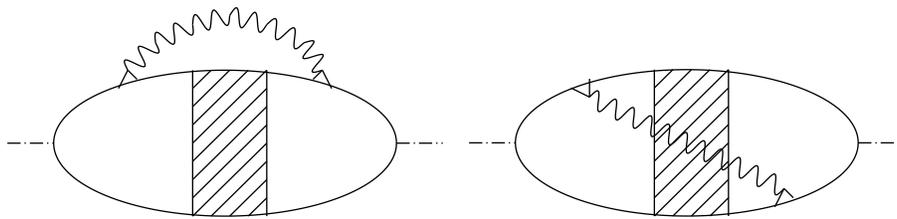
\end{document}